\documentstyle[buckow,graphicx]{article}

\def\Tr{\rm{Tr}\;}

\def\tr{\rm{tr}\;}

\newcommand{\text}[1]{\rm{#1}}

\newcommand{\eqn}[1]{(\ref{#1})}

\def\beq{\begin{equation}}
\def\eeq{\end{equation}}

\begin {document}


\def\titleline{Getting the D-brane effective action from
BPS configurations}

\def\authors{
P.\ Koerber\footnote{Aspirant FWO} and A.\ Sevrin }
\def\addresses{
Theoretische Natuurkunde - Vrije Universiteit Brussel \\
Pleinlaan 2, B-1050 Brussels, Belgium\\
{\tt koerber, asevrin@tena4.vub.ac.be}
}

\def\abstracttext{
We review a method to find the non-abelian open superstring effective action, 
thereby settling the issue of the ordering ambiguities.
We start from solutions to Yang-Mills which, in D-brane context, define certain BPS configurations.
Studying their deformations in the abelian case shows that the Born-Infeld action is the
{\em unique} deformation which admits solutions of this type. By applying the method to the non-abelian
case we calculated the full effective action through $ {\cal O}(\alpha'{}^3)$. The presence of derivative 
terms turns out to be essential. Testing the result by comparing the spectrum in the presence of a 
constant magnetic background field with the string theory prediction, we obtain perfect agreement.
}

\large

\makefront

\section{Introduction}

A most attractive feature of D-branes is their close connection to gauge theories. 
Indeed the massless bosonic worldvolume degrees of freedom of a D$p$-brane consist of a $U(n)$
gauge field and $9-p$ $U(n)$-valued scalar fields. In leading order in $\alpha'$ the 
worldvolume action is precisely the supersymmetric $U(n)$ Yang-Mills theory dimensionally reduced to $p+1$
dimensions \cite{witten}. In the remainder of this paper, we will ignore both the fermions and the scalar fields.  
When $n=1$, the abelian case, the full effective action is known for constant fieldstrengths: it is 
the Born-Infeld action \cite{ts}. For the non-abelian case, $n\geq 2$, no such result is known. 

A direct calculation requires matching the effective action to $N$-point open
superstring amplitudes. This has been done for $N\leq 4$, yielding the full effective
action through order $ {\cal O}(\alpha'{}^2)$ \cite{direct}, \cite{direct1}, \cite{bilal} and derivative terms
at higher orders \cite{bilal2}. Pushing the direct calculation to higher orders seems presently infeasible.

In this paper, we review a powerful method which allows for an indirect calculation
of the full effective action, including derivative terms, order by order in $\alpha'$ \cite{abelian}. 
Stable holomorphic bundles 
define solutions to Yang-Mills which generalize the standard notion of instantons to arbitrary dimensions.
In D-brane context, such solutions correspond to BPS configurations in the weak field limit. Requiring that
these solutions, or some deformation thereof, solve the equations of motion of the full effective action allows
one to determine both the equations of motion and the deformation of the solution order by order 
in $\alpha'$ .
This program was carried out through $ {\cal O}( \alpha'{}^3)$ in \cite{nonabelian} and tested in \cite{spectrum}.

\section{BPS states of D-branes}

Simple BPS configurations of D-branes arise as follows \cite{angles}.  
Start with two coinciding Dp-branes in the $(1,3,\ldots,2p-1)$ directions.
Keeping one of them fixed, rotate the other one 
over angles $\phi_i$ in the (2i-1\,2i) plane, for $1\leq i\leq p$.
When
\begin{eqnarray}
\sum_{i=1}^p \phi_i & = 2 \pi n, \label{BPScond}
\end{eqnarray}
holds, one finds that $32/2^p$ supersymmetries are 
preserved\footnote{Note that it is possible to preserve more supersymmetries with more
stringent BPS conditions.}. 

Next, we T-dualize the system in the $2,4,\cdots,p$ directions,
ending up with
two coinciding D$2p$-branes with magnetic fields turned on. Indeed, having
two D$2p$-branes extended in the 1, 2, ..., 2p directions with constant magnetic
flux $F_{2i-1\, 2i}=f_i\sigma_3$, $i\in\{1,\cdots,p\}$, and all other components zero,
we can choose a gauge in which the potentials are given by,
\begin{eqnarray}
A_{2i-1}=0,\quad A_{2i}=f_i\, x^{2i-1}\,\sigma_3.\label{pot}
\end{eqnarray}
T-dualizing back, one ends up with two D$p$-branes
with transversal coordinates given by
\begin{eqnarray}
X^{2i}=2\pi\alpha 'A_{2i}.\label{trc}
\end{eqnarray}
Using eq.\ (\ref{pot}) in eq.\ (\ref{trc}), we recognize the original
configuration with the two D$p$-branes at angles with the angles given by
\begin{eqnarray}
\phi_i&=& 2\arctan(2\pi\alpha ' f_i) \, .
\end{eqnarray}
In terms of the magnetic field, the BPS condition \eqn{BPScond} is formulated as,
\beq
\sum_i 2 \arctan (2\pi \alpha' f_i) = 2\pi n \, ,
\label{BPSf}
\eeq
or in the limit of weak fields, $\alpha' F \rightarrow 0$,
\beq
\sum_i f_i = 0 \, .
\label{BPSweak}
\eeq
Since a BPS configuration should solve the equations of motion, eq.\ \eqn{BPSweak} 
should provide solutions to the Yang-Mills equations of motion.

Switching to complex coordinates, $z^{\alpha}=(x^{2\alpha-1}+i x^{2\alpha})/\sqrt{2}$,
$\bar z^{\bar \alpha}=(z^\alpha)^*$, one finds that gauge field configurations satisfying
\beq
\sum_{\alpha} F_{\alpha\bar{\alpha}} = 0 \, .
\label{DUY}
\eeq
\beq
F_{\alpha\beta} = 0 \, , \quad F_{\bar{\alpha}\bar{\beta}} = 0 \, .
\label{holo}
\eeq
solve the Yang-Mills equations of motion,
\beq
D_{\bar{\alpha}} F_{\alpha\bar{\beta}} + D_{\alpha} F_{\bar{\alpha}\beta} = 
D_{\bar{\beta}} F_{\alpha\bar{\alpha}} = 0 \, ,
\eeq
where we used the Bianchi identities. Furthermore, using the supersymmetry
transformation rule, $\delta\psi \propto  F_{\mu\nu}\gamma^{\mu\nu}\epsilon$,
one discovers that these configurations do preserve $32/2^p$ supersymmetries.
The configurations studied in \eqn{BPSweak} are a special case: the fieldstrengths are constant
and in addition, $F_{\alpha\bar\beta}=0$ for $\alpha\neq\beta$. Eq.\ (\ref{DUY}) reduces then to eq.\ (\ref{BPSweak}).

Solutions satisfying eqs.\ (\ref{DUY}) and (\ref{holo}) define a {\em stable holomorphic vector bundle}.
For $p=2$ ($d=4$) they are equivalent to the standard instanton equations.

\section{Abelian Born-Infeld}

The next step is to look for deformations of the Yang-Mills action 
which still allow for this type of solutions.
We work in the slowly varying field limit, so that we do not have to consider derivative terms.
We expect condition \eqn{holo} not to get any corrections because of its geometric origin.
Condition \eqn{DUY} on the other hand {\em will} get corrections.

As an illustration we will consider the calculation through order $\alpha'^2$.  
The most general lagrangian\footnote{Order $\alpha'$ is zero in the abelian case because of
antisymmetry.} through this order reads\footnote{From now on we put $2 \pi \alpha'=1$.},
\beq
\frac{1}{4} \tr F^2 + \lambda_{(0,1)}(\tr F^4)+\lambda_{(2)}(\tr F^2)^2 + {\cal O}(\alpha'^{4}) \, ,
\eeq
where tr denotes tracing over the Lorentz-indices and we understand that repeated indices are
summed over in what follows.
This leads to the following equations of motions --- in complex coordinates and using eq.\ \eqn{holo},
\beq
\partial_{\bar{\alpha}} F_{\alpha\bar{\beta}} + 8 \lambda_{(0,1)} \partial_{\bar{\alpha}} F^3_{\alpha\bar{\beta}} +
16 \lambda_{(2)} \partial_{\bar{\alpha}} (F^2_{\gamma\bar{\gamma}} F_{\alpha\bar{\beta}}) =0\, ,
\eeq
where:
\beq
F^l_{\alpha\bar{\beta}} = F_{\alpha\bar{\alpha_2}} F_{\alpha_2\bar{\alpha_3}} \ldots F_{\alpha_{l}\bar{\beta}} \, .
\eeq
After applying Bianchi identities, this leads to:
\begin{eqnarray}
& \partial_{\bar{\beta}} \left( F_{\alpha\bar{\alpha}} + \frac{8 \lambda_{(0,1)}}{3}F^3_{\alpha\bar{\alpha}}\right) & \text{deformation \, condition \, \eqn{DUY}} \nonumber \\
& + \left( \frac{8 \lambda_{(0,1)}}{2} + 16 \lambda_{(2)}\right)\partial_{\bar{\gamma}} F^2_{\alpha\bar{\alpha}} F_{\gamma\bar{\beta}} & \text{condition \, between \, \lambda_{(0,1)} \, and \, \lambda_{(2)}} \nonumber \\
& + 16 \lambda_{(2)} \partial_{\bar{\beta}}F_{\alpha\bar{\alpha}} F^2_{\gamma\bar{\gamma}} + 8 \lambda_{(0,1)} \partial_{\bar{\gamma}}F_{\alpha\bar{\alpha}} F^2_{\gamma\bar{\beta}} = 0 & \text{``1-loop \, terms"} \, .
\label{eee}\end{eqnarray}
The first line vanishes, provided we add an $ {\cal O} (\alpha'{}^2)$ correction to eq.\ (\ref{DUY}).
The second line vanishes when imposing $\lambda_{(0,1)}+4\lambda_{(2)}=0$. 
The last line vanishes by virtue of eq.\ (\ref{DUY}). Note that because we introduced an 
$ {\cal O} (\alpha'{}^2)$ correction to eq.\ (\ref{DUY}), the last line in
eq.\ (\ref{eee}) will contribute to the equations of motion at
$ {\cal O} (\alpha'{}^4)$. This will 
eventually lead to conditions between coefficients at different orders in $\alpha'$.

For general terms in the lagrangian:
\beq
\lambda_{(p_1,\ldots,p_n)}(\tr F^2)^{p_1}\ldots(\tr F^{2n})^{p_n} \, ,
\eeq
we find by continuing the same procedure:
\beq
\lambda_{(p_1,\ldots,p_n)}=\frac{(-1)^{p_1+\cdots+p_n+1}}{4^{p_1+\cdots+p_n}}\frac{1}{p_1!\ldots p_n!}\frac{1}{1^{p_1}\ldots n^{p_n}} \,
\eeq
which corresponds to the expansion 
of the abelian Born-Infeld action $-\sqrt{\det\left(\delta^{\mu}_{\nu}+F^{\mu}_{\nu}\right)}$.
The deformed stability condition is
\beq
F_{\alpha\bar{\alpha}}+\frac{1}{3}F^3_{\alpha\bar{\alpha}} + \frac{1}{5}F^5_{\alpha\bar{\alpha}}+ \cdots = 0 \, ,
\eeq
or using $F_{\alpha\bar{\alpha}} = i F_{x^{2\alpha-1}x^{2\alpha}}$ exactly condition \eqn{BPSf} for $n=0$.

We therefore conclude that the abelian Born-Infeld action is the unique deformation of Yang-Mills (without derivative terms) that admits
generalized stable holomorphic vector bundles as a solution.  

\section{From the abelian to the non-abelian case}

\begin{figure}[tbp]
\centering
\includegraphics[scale=0.5]{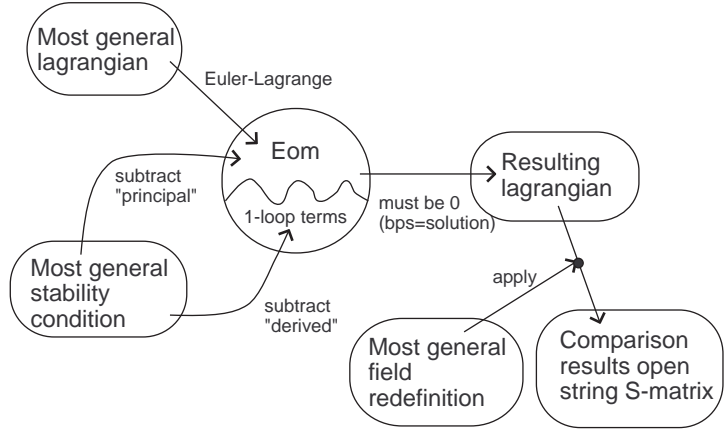}
\caption{A very rough flowchart of the calculations in the non-abelian case.}
\label{flowchart}
\end{figure}

The same method works essentially for the non-abelian case too, but there are several complications:
\begin{itemize}
\item Because of identities of the form $[D_{\mu},D_{\nu}]F_{\rho\sigma}=[F_{\mu\nu},F_{\rho\sigma}]$, 
there is no unambiguous notion
of slowly varying fieldstrengths as in the abelian case.  
In fact, our method clearly shows that from order $ \alpha'{}^4$ on, derivative terms are unavoidable.
\item Once derivative terms are included, field redefinition ambiguities have to be dealt with! 
This fact should also be taken into account when comparing to results in the literature.
\item In the non-abelian case, there is a huge amount of possible terms 
(derivative terms, permutations of the $F$s), that are connected
by a complex web of identities (partial integration identities, Bianchi identities and $[D,D]F=[F,F]$-identities, 
field redefinitions). So we wrote a computer program to keep track of all these.
\end{itemize}
A (very rough) flowchart of the calculations would look as in figure \ref{flowchart}.

In this way we found up to order $\alpha'^3$ and modulo field redefinitions:
\begin{eqnarray}
{\cal L}&=&\frac{1}{g^2}
\Tr\Big(-\frac 1 4
F_{\mu_1}{}^{\mu_2}F_{\mu_2}{}^{\mu_1} -
\frac{1}{24} F_{\mu_1}{}^{\mu_2}F_{\mu_2}{}^{\mu_3}F_{\mu_3}{}^{\mu_4}F_{\mu_4}{}^{\mu_1} -
\frac{1}{12}F_{\mu_1}{}^{\mu_2}F_{\mu_2}{}^{\mu_3}F_{\mu_4}{}^{\mu_1}F_{\mu_3}{}^{\mu_4}+
 \nonumber \\
&& \frac{1}{48}F_{\mu_1}{}^{\mu_2}F_{\mu_2}{}^{\mu_1}F_{\mu_3}{}^{\mu_4}F_{\mu_4}{}^{\mu_3}+
\frac{1}{96}F_{\mu_1}{}^{\mu_2}F_{\mu_3}{}^{\mu_4}F_{\mu_2}{}^{\mu_1}F_{\mu_4}{}^{\mu_3}-
\nonumber\\
&&\Lambda\,\Big(
F_{\mu_1}{}^{\mu_2}F_{\mu_2}{}^{\mu_3}F_{\mu_3}{}^{\mu_4}F_{\mu_5}{}^{\mu_1}F_{\mu_4}{}^{\mu_5}+
F_{\mu_1}{}^{\mu_2}F_{\mu_4}{}^{\mu_5}F_{\mu_2}{}^{\mu_3}F_{\mu_5}{}^{\mu_1}F_{\mu_3}{}^{\mu_4}-
\nonumber\\
&&\frac 1 2
F_{\mu_1}{}^{\mu_2}F_{\mu_2}{}^{\mu_3}F_{\mu_4}{}^{\mu_5}F_{\mu_3}{}^{\mu_1}F_{\mu_5}{}^{\mu_4}
+F_{\mu_1}{}^{\mu_2}\left(D^{\mu_1}F_{\mu_3}{}^{\mu_4}\right)\left(D_{\mu_5}
F_{\mu_2}{}^{\mu_3}\right)F_{\mu_4}{}^{\mu_5}-
\nonumber\\
&&\frac 1 2
\left(D^{\mu_1}F_{\mu_2}{}^{\mu_3}\right)\left(D_{\mu_1}F_{\mu_3}{}^{\mu_4}\right)
F_{\mu_5}{}^{\mu_2}F_{\mu_4}{}^{\mu_5} -\frac 1 2
\left(D^{\mu_1}F_{\mu_2}{}^{\mu_3}\right)F_{\mu_5}{}^{\mu_2}\left(D_{\mu_1}
F_{\mu_3}{}^{\mu_4}\right)F_{\mu_4}{}^{\mu_5} +\nonumber \\
&&\frac 1 8
\left(D^{\mu_1}F_{\mu_2}{}^{\mu_3}\right)F_{\mu_4}{}^{\mu_5}\left(D_{\mu_1}
F_{\mu_3}{}^{\mu_2}\right)F_{\mu_5}{}^{\mu_4} -
\left(D_{\mu_5}F_{\mu_1}{}^{\mu_2}\right)F_{\mu_3}{}^{\mu_4}\left(D^{\mu_1}
F_{\mu_2}{}^{\mu_3}\right)F_{\mu_4}{}^{\mu_5}\Big)\Big)
, \nonumber\\\label{order3}
\end{eqnarray}
where $\Tr$ is the group trace. $\Lambda$ is an arbitrary constant which can be
fixed by comparing to string scattering calculations, see for instance \cite{bilal2}:
\beq
\Lambda = - \frac{2 \zeta(3)}{\pi^3}.
\eeq
String theory tells us that the order $ \alpha'{}^m$ correction to the effective action is proportional to
$\zeta(m)/\pi^m$. Since Euler, we know that for $m$ even this is rational while for $m$ odd it is not.
Our method clearly only yields rational numbers. So it is most fortunate that we obtained a free parameter at
order $ \alpha'{}^3$. In fact we expect our method to fully fix the action at even orders but leaving free parameters at
odd orders.

\section{Checks on the result}

\begin{itemize}
\item Internal check: before finding eq.\ \eqn{order3} our program had to solve a set of 156 homogeneous equations
in 63 unknowns.  The fact that we found a solution is encouraging!
\item Fluctuation spectrum: string theory predicts the following spectrum for strings
stretching between the two $Dp$-branes
at angles from section 2:
\beq
M^2=\left(\sum_j (2 n_j + 1)\phi_j \right) \pm 2 \phi_i \, ,
\eeq
with $\phi_i=2 \arctan f_i$.

Fluctuations around Yang-Mills (order $\alpha'^0$) on the $2p$-torus lead to the following spectrum:
\beq
M^2=\left(\sum_j 2(2 n_j + 1) f_j \right) \pm 4 f_i \, ,
\eeq
and the higher orders in the $f_j$ must come from higher order terms in $\alpha'$.  We see immediately
that the odd orders should not contribute.

Careful calculations \cite{spectrum} reveal that indeed our order $\alpha'^3$ does not contribute to the spectrum.
\end{itemize}

\section{Conclusions and future research}

Because of these checks, we are fairly confident that the result is indeed the correct non-abelian open
superstring effective action through ${\cal O}(\alpha'^3)$.

Furthermore our program managed to calculate the lagrangian through ${\cal O}(\alpha'^4)$. Unfortunately, at this order not only the calculation but
also the result is very complicated.  Due to the web of identities, and especially the field redefinitions, we lack a sort of
``canonical'' form.  This is probably not the way to get all order results in $\alpha'$.

Therefore, we plan to make an expansion in the degree of non-abelianality, where the zeroth order would be the symmetrized trace Born-Infeld
action.  Using our method, we try to calculate corrections --- at the first non trivial order in this degree and at all
orders in $\alpha'$.

Finally, let us make a remark about the D3-brane effective action versus the $d=4$, $N=4$ effective super
Yang-Mills action. In the abelian case, the $ F^8$ term in the one-loop $N=4$ super
Yang-Mills effective action is different in structure, \cite{buch},
from the $F^8$ term in the Born-infeld action \cite{ft}.
In the non-abelian case, this discrepancy shows up at lower order. Indeed, in \cite{milaan},
the terms of the same dimensions as the ones discussed in this paper ($F^5$ and $D^2F^4$),
in the one-loop effective action of $N=4$
supersymmetric Yang-Mills in four dimensions were calculated. Not only are these terms
different from the ones calculated in this paper \cite{nonabelian}, 
they do not pass the test in \cite{spectrum} as well.
\vspace{.3cm}

{\bf Acknowledgments}
P.K. and A.S. are
supported in part by the
FWO-Vlaanderen and in part by the European Commission RTN programme
HPRN-CT-2000-00131, in which the authors are associated to the university
of Leuven.



\end{document}